\documentclass{hep99} 
\usepackage{float,flafter,epsfig}

\def\be{\begin{equation}}
\def\ee{\end{equation}}
\def\br{\begin{eqnarray}}
\def\er{\end{eqnarray}}
\def\NPB#1#2#3{{\it Nucl.~Phys.} {\bf{B#1}} (19#2) #3}
\def\PLB#1#2#3{{\it Phys.~Lett.} {\bf{B#1}} (19#2) #3}
\def\PRD#1#2#3{{\it Phys.~Rev.} {\bf{D#1}} (19#2) #3}
\def\PRL#1#2#3{{\it Phys.~Rev.~Lett.} {\bf{#1}} (19#2) #3}

\begin{document}
\begin{titlepage}

\Large

\begin{center}
{\bf \LARGE Domain walls in supersymmetric QCD} \\
\vspace{1.5cm}
B. de Carlos\footnote{\Large e-mail address: Beatriz.de.Carlos@cern.ch} \\
{\em Theory Division, CERN, CH-1211 Geneva 23, Switzerland} \\
\vspace{0.5cm} 
J.M. Moreno\footnote{\Large  e-mail address: jesus@makoki.iem.csic.es} \\
{\em Instituto de Estructura de la Materia, CSIC}  \\
{\em Serrano 123, 28006 Madrid, Spain}
\end{center}
\vspace{1.7cm}
\begin{abstract}
\noindent
We consider domain walls that appear in supersymmetric $SU(N)$ 
with one massive flavour. In particular, for $ N \geq 3 $ we 
explicitly construct the elementary domain wall that interpolates 
between two contiguous vacua.  We show that these solutions are 
BPS saturated for any value of the mass of the matter fields. 
We also comment on their large N limit and their relevance for 
supersymmetric gluodynamics.
\end{abstract}

\vspace{1.7cm}

\begin{center}
{Talk given at the  {\bf EPS Conference on High Energy Physics} 
Tampere, Finland, July 1999} \\
{(\em to be published in the proceedings)}\\
\end{center}

\thispagestyle{empty}

\vspace{2.5cm} 
\noindent
CERN-TH/99-322 \\
IEM-FT-197/99 \\
IFT-UAM/CSIC-99-41 \\
October 1999

\vskip-22cm
\rightline{CERN-TH/99-322}
\vskip3in

\end{titlepage}

\title{Domain walls in supersymmetric QCD}

\author{B. de Carlos$^1$ and J.M. Moreno$^2$}

\address{
$^1$ Theory Division, CERN, CH-1211 Geneva 23, Switzerland \\
$^2$ Instituto de Estructura de la Materia, CSIC, Serrano 123, 
28006 Madrid, Spain\\[3pt]
 E-mails: {\tt  Beatriz.de.Carlos@cern.ch, jesus@makoki.iem.csic.es }}

\abstract{
We consider domain walls that appear in supersymmetric $SU(N)$ 
with one massive flavour. In particular, for $ N \geq 3 $ we 
explicitly construct the elementary domain wall that interpolates 
between two contiguous vacua.  We show that these solutions are 
BPS saturated for any value of the mass of the matter fields. 
We also comment on their large N limit and their relevance for 
supersymmetric gluodynamics.
}

\maketitle


The study of supersymmetric gauge theories that are in the strong coupling 
regime has been intensified recently by the realization that some of
their constructions could admit exact solutions. In particular, the issue
of domain walls in SU(N) supersymmetric gluodynamics, the theory of gluons 
and gluinos, is one of the most exciting ones. These walls arise 
because this theory has an axial U(1) symmetry broken by the anomaly to a 
discrete $Z_{2N}$ chiral symmetry. Due to non-perturbative effects gluino 
condensates $\left( \langle \lambda \lambda \rangle \right)$ form, breaking 
the symmetry further down to $Z_2$. This leaves us with a set of $N$ 
different vacua labelled by
\be
\langle {\rm Tr} \lambda \lambda \rangle = \Lambda^3 e^{2 \pi i k/N} \;\; 
k= 0,1,...,N-1 \;,
\ee
where $\Lambda$ is the condensation scale, and, as indicated above, a set
of domain walls interpolating between them. If we assume that they are BPS 
saturated, the energy density of these walls is exactly calculable and 
given by \cite{Dvali97,Kovner97,Chibisov98}
\be
\label{tension}
\epsilon = \frac{N}{8 \pi^2} | \langle {\rm Tr} \lambda \lambda 
\rangle_{\infty} - \langle {\rm Tr} \lambda \lambda \rangle_{-\infty}| \;\;,
\ee
In fact, it has been shown in Ref.~\cite{Dvali99} that, in the large 
$N$ limit, these domain walls are BPS states. On the other hand, these 
solutions preserving half of the supersymmetry would play an important role 
in the D-brane description of $N=1$ SQCD~\cite{Witten97}.

In order to get to study pure gluodynamics, it is convenient to add
matter fields to the theory and analyze the limit where these extra 
fields (usually taken to be pairs of chiral superfields transforming 
as $(N, {\bar N})$ under the colour group) become very heavy. In the 
strong coupling regime, we should expect the formation of squark 
condensates. These models were considered in Refs~\cite{Smilga,SmilgaSUN}, 
for the case of ($N-1$) flavours. Their analysis of the vaccuum structure 
led to the conclusion that the existence of BPS saturated domain 
walls was restricted to values for the mass $m$ of the squark fields 
below a certain critical one. It therefore looked like it would be
impossible to recover pure gluodynamics by taking the limit 
$m \rightarrow \infty$, and this is precisely the main issue we will
address during this talk.

In order to do that let us consider supersymmetric QCD with $SU(N)$ gauge 
group and one couple of chiral superfields $(Q,\bar{Q})$ 
transforming as $N,\bar{N}$. Non-perturbative effects become relevant at 
the scale $\Lambda$, where condensates form. The gaugino and squark 
colourless condensates are described by the following composite fields
\br
S & = & \frac{3}{32 \pi^2} {\rm Tr} W^2 \;\;, \nonumber \\
\\ 
M & = & Q \bar{Q} \;\;, \nonumber
\er
where $W^2$ is the composite chiral superfield whose lowest component is
$\lambda \lambda$. In this regime, the relevant degrees of freedom are 
described by a Wess-Zumino model, as shown in Ref.~\cite{Venez82}. Its 
effective Langrangian is given by
\be
{\cal L} = \frac{1}{4} \int d^4 \theta \; {\cal K} + \frac{1}{2} \left[ 
\int d^2 \theta \; {\cal W} + {\rm h.c.} \right] \;\;,
\ee
where ${\cal K}$ is the K\"ahler potential and ${\cal W}$ is the 
superpotential
\be
{\cal W} = \frac {2}{3} S
\ln \frac{S^{N-1}  M}{ \Lambda^{3 N - 1} e^{N-1}} -
\frac {1}{2}  m M \;\; ,
\label{TVY1}
\ee
with $m$ the mass for the matter superfields. 
This superpotential has $N$ extrema labeled by the different phases of 
the gaugino condensate. At the minimum we have the gaugino condensate 
fixed to
\be
S_*^{N} = \frac {3}{4}   m \Lambda^{3 N - 1}\;\;.
\label{gluino}
\ee
The matter condensate is aligned with respect to the former and given by
\be
M_*  =   \frac {1}{m} \frac {4}{3}  S_* \;\;.
\label{matter}
\ee
We want to study domain wall configurations that interpolate between the 
different minima. Here a technical problem appears: the superpotential has 
several branches associated with its logarithmic piece~\cite{Kogan98}. In 
the pure SUSY gluodynamics limit described by Veneziano and 
Yankielowicz~\cite{Venez82} this is a severe problem, since any 
configuration connecting two vacua has to cross this branch. 
This is not necessarily the case when we include other fields, given that 
the variation in the phase of the gaugino condensate can be partially 
compensated by these new fields. In this case, this will be done by matter 
fields.

Let $(S,M)_a$ be a particular vacuum. We can continuously deform it into 
another vacuum, $(S,M)_b$. For this path in the configuration space, we
define $\delta$, $w$ such that
\br
S|_b     & = & e^{ i \delta } S |_a\;, \nonumber \\
\label{transf} \\
M |_b & = & e^{ i (\delta + 2 \pi w )}  M |_a \;.   \nonumber
\er
Notice that Eq. ~(\ref{matter}) implies that
$w$ must be some integer number. 
On the other hand, one necessary condition to avoid crossing the 
logarithmic branch along this path is
\be 
(N-1) \delta +  (\delta + 2 \pi w) = 0 \;.
\ee
Since we are interested in configurations interpolating between the 
vacua $i$ and $i-1$, we will fix  $\delta =  -\frac{2 \pi}{N}$ and then
$ w =  1 $.

If we assume that there is a BPS domain wall connecting these two
vacua, it will be described by the following 
differential equations
\br
{\cal K}_{S{\bar S}} \partial_z {\bar S}     & = & e^{i\gamma}  
\frac{ \partial{{\cal W}} } {\partial S } \;\; , \nonumber \\
\label{BPSeqs} \\
{\cal K}_{M{\bar M}} \partial_z {\bar M_i^i} & = & e^{i\gamma}  
\frac{ \partial{{\cal W}} } {\partial M^i_i } \;\;, \nonumber
\er
where ${\cal K}_{\phi{\bar \phi }}$ is the induced metric from the K\"ahler 
potential ${\cal K}$, and $\gamma$ is given by
\be
\gamma = -\frac{1}{2}(\delta + \pi) =  \frac{\pi}{ N} -\frac{\pi}{2}
\;\;.
\ee
The configuration is described by four real functions 
\br
M (z) & = & |M_*|  \; \rho(z)  e^{i\alpha (z) }  \;\;, \nonumber \\
\label{conds} \\
S (z) & = & |S_*| \;  R(z) e^{i\beta (z) } \;\;.  \nonumber
\er
Notice that we have defined $\rho(z), R(z)$ in such a way that 
$\rho(\pm \infty) = R (\pm \infty) = 1 $. On the other hand, $\alpha$ varies 
from $0$ to $2\pi(1-1/N)$ and $\beta$ from $0$ to  
$-2 \pi /N$. A consistent ansatz under reflection 
$z\rightarrow -z$ is given by: $\rho(z)=\rho(-z)$, $R(z)=R(-z)$, 
$\beta(z) = - 2 \pi/N  - \beta(-z)$ and
$\alpha(z) =  2 \pi (1- 1/N)   - \alpha(-z)$.
These relations fix the boundary conditions at $z=0$.
Eqs~(\ref{BPSeqs}) imply the following BPS constraint
\be
Im \left[ e^{i \gamma} {\cal W} (S, M) \right] = const \;\;.
\ee

In Refs~\cite{Smilga,SmilgaSUN} a similar analysis was presented for 
$N_f = N-1$ couples of matter fields. In these papers it was shown 
that these domain walls are BPS states only for squark masses lower than 
some critical value, $m_*$, that depends on $N$ and the K\"ahler 
potential. The existence of this bound is related to the presence of two 
different BPS domain wall solutions for small enough values of $m$, which 
became identical at the critical value.

\begin{figure}
\centerline{  
\psfig{figure=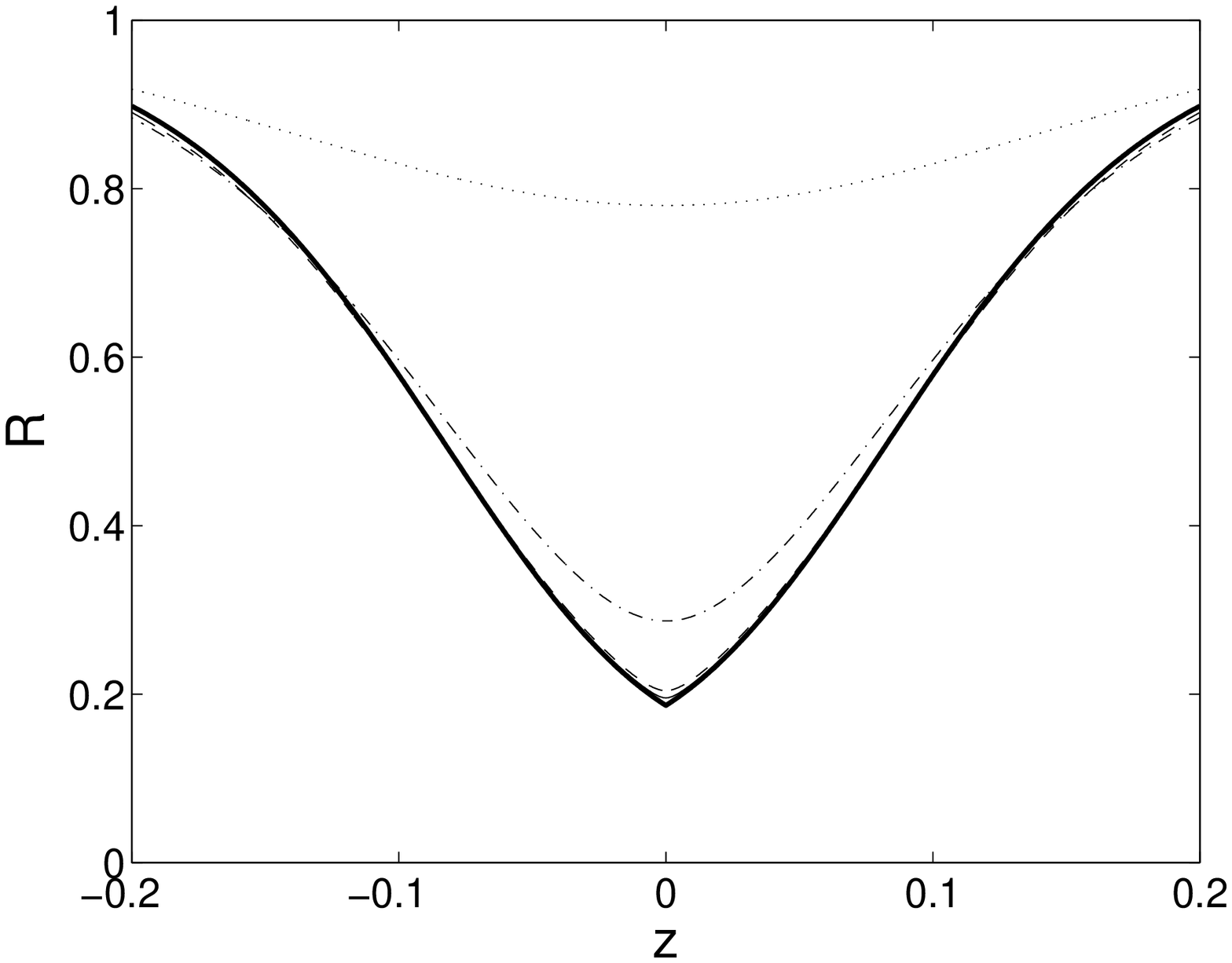,height=7cm,width=9cm,bbllx=0cm,bblly=7cm,bburx=21cm,bbury=21cm}
}
\caption{}
{\footnotesize \noindent $R(z)$ as defined in Eqs~(\ref{conds}) versus $z$ 
(in units of ${\tilde{\Lambda}}^{-1}$), for $m = 2$ (dotted line), $20$ 
(dash-dotted), $100$ (dashed), $200$ (solid). The thick solid line corresponds
to the $m \rightarrow \infty$ solution given by Eq.~(\ref{limit}).}
\label{fig2}
\end{figure}

Here we have worked in detail a different case: $N=3$, $N_f=1$,
using the same  K\"ahler potential, i.e. 
${\cal K} =  (S {\bar S})^{1/3}  + (M {\bar M})^{1/2} $.
We have found that the 
equations can be solved for {\em all} values of the squark mass, and we have 
checked that the logarithmic branch is never crossed. The 
profiles for $R$ are shown in Fig.~\ref{fig2} for 
several values of $m$ (given in units of $\Lambda$), focusing on their 
central region. The spatial coordinate $z$ is expressed in units of 
${\tilde \Lambda}^{-1}$,  where ${\tilde \Lambda} = \Lambda  
(\frac{3 m}{4\Lambda})^{1/3N} $ is the effective QCD scale that arises 
in the large $m$ limit.

In our case there is only one BPS solution for every value of $m$. 
This can be understood analyzing both the large and small $m$ limit,
as explained in detail in \cite{dCM99}. These limits depend on the
number of flavours.

Let us consider the limit $m \rightarrow \infty$, that is expected to 
describe pure gluodynamics. From Fig.~\ref{fig2} we see that there is a 
well defined gaugino condensate profile in that limit. In fact, the 
following constraints apply in the asymptotic regions
\br
\rho (z) e^{i\alpha(z)}  & = &   R(z) e^{i  \beta(z)} \; \;   
(z << -1 /m) \;\;, \nonumber   \\
\label{inf} \\
\rho (z) e^{i\alpha(z)}  & = &   R(z) e^{i (\beta(z) - 2 \pi)}  \; \;   
(z >> 1 /m) \;\;.  \nonumber
\er
Therefore, using Eq.~(\ref{inf}) in this large $m$ limit we can get rid of 
$\alpha$ and $\rho$ in the BPS equations. Also the BPS constraint involves 
only the gaugino condensate and can be written as
\be
Im \left\{ e^{i \left[ \gamma + \beta(z) \right]} R(z)  
\left[ \ln \left( R(z) e^{i \tilde{\beta}(z)} \right) -1 \right]
\right\} =  const \;\;,  
\label{split}
\ee
where $\tilde{\beta}(z) =  \beta(z)$ for $z<0$ and $\tilde{\beta}(z) =  
\beta(z) + 2\pi/N$ for  $z>0$. This constraint allows us to 
express $\beta$ as a function of $R$, and we end up with the following 
BPS equation for $R(z)$
$$
\partial_z R (z)  =  6 N (R(z))^{4/3} \tilde{\Lambda}  
\left\{ \cos (\gamma +  \beta ([R(z)]) )\; \ln R(z)  \right. 
\nonumber 
$$
\be
 -  \left. \sin (\gamma + \beta( [R(z)])) \; \tilde{\beta} [R(z)] 
\right\} \;\;,
\label{limit}
\ee
Notice that, in the large $m$ limit, the profile has two branches
which are smoothly connected for finite values of $m$. In this limit, 
the domain wall has two typical scales. One is associated to the gaugino 
condensate, while the other one is associated to the additional field that
compensates the phase changes of the gaugino condensate, in our case the 
matter field\footnote{See \cite{Dvali99} for other possibilities.}.

Let us now analyze the large N limit of this asymptotic 
configuration \cite{us}. In order to do that we expand the normalized 
modulus
\be
R(z) = 1 -r(z)/N + {\cal O}(1/N^2)\;\;,
\ee
and we rewrite $ \beta (z) = b(z) \frac{2 \pi}{N}$. Then 
Eq.~(\ref{split}) implies
\be
b(z) = - r(z) 
\ee
for the left branch, i.e. for $z<0$. If we want to analyze the structure 
of the domain wall, we have to include properly the dependence of $S$ on 
$N$. One expects \cite{Dvali99,Witten97}
\be
S \equiv \langle {\rm Tr} \lambda_\alpha  \lambda^\alpha \rangle \sim 
{\cal O} (N) \;\;.
\ee
Then, we have to replace $\Lambda^3$ by $N \Lambda ^3$ in the above
equations. This implies that the energy of the previous elementary
domain wall, as given by Eq.~(\ref{tension}), scales like
$\epsilon \sim {\cal O}(N)$. If we want to study the width, i.e. 
how the  energy is spread along the domain wall, we need some
information about the $N$ dependence of kinetic terms of both gaugino
and regulator -matter- fields.

For example, if we assume that the gaugino kinetic term
term is ${\cal O}(N^a)$, then $r(z)$ is given by
\be
\partial_{|z|}  r(|z|)  =   - k^2 \tilde{\Lambda} N^{1-a} r(|z|)  \;\;,
\ee
where the precise value of the positive constant $k^2$ depends on the
particular  K\"ahler potential. Under this assumption, the scale
associated with the gaugino field variations is 
${\cal O} ( 1/ (\tilde{\Lambda} N^{1-a})) $.

In summary, it is possible to build BPS domain walls in SQCD with one 
flavour for any value of the mass of the matter fields. This allows us 
to study the limit where the theory approaches pure supersymmetric 
gluodynamics.

\vspace{0.2cm}

The work of JMM was supported by CICYT of Spain (contract AEN98-0816).
We both acknowledge the British Council/Acciones Integradas program for the 
financial support received through the grant HB1997-0073.

\end{document}